# Flow Dimension and Capacity for Structuring Urban Street Networks

Bin Jiang

Department of Land Surveying and Geo-informatics, The Hong Kong Polytechnic University
Hung Hom, Kowloon, Hong Kong, Email: bin.jiang@polyu.edu.hk

**Abstract:** This paper aims to measure the efficiency of urban street networks (a kind of complex networks) from the perspective of the multidimensional chain of connectivity (or flow). More specifically, we define two quantities: flow dimension and flow capacity, to characterize structures of urban street networks. To our surprise for the topologies of urban street networks, previously confirmed as a form of small world and scale-free networks, we find that (1) the range of their flow dimension is rather wider than their random and regular counterparts, (2) their flow dimension shows a power-law distribution, and (3) they have a higher flow capacity than their random and regular counterparts. The findings confirm that (1) both the wider range of flow dimension and the higher flow capacity can be a signature of small world networks, and (2) the flow capacity can be an alternative quantity for measuring the efficiency of networks or that of the individual nodes. The findings are illustrated using three urban street networks (two in the Europe and one in the USA).

## 1. Introduction

Urban street networks can be seen as an interconnected graph from two different and contrast perspectives. On the one hand, an urban street network is regarded as a graph with nodes and edges representing respectively the junctions and street segments (between two adjacent junctions). This can be called a geometric approach. On the other hand, in contrast to the geometric approach, a topological approach regards an urban street network as an interconnected graph consisting of nodes representing individual named streets (Jiang and Claramunt 2004), and edges linking the nodes if the corresponding streets are intersected. The difference between the two approaches is subject to whether or not geometric distance plays an important role in the graph theoretical representations. For the geometric approach, the graph edges (street segments) are weighted by the distance between the two corresponding junctions. With the topological approach, an entire street is collapsed as a node, and all the edges have a unit weight (Porta et al. 2006). In other words, there is no weighting difference among the edges. The geometric approach is an advantageous representation in conventional network analysis (Haggett and Chorley 1969) and transportation modelling (Miller and Shaw 2001) with geographic information systems (GIS), but it is not good for illustrating the structure of urban street networks, since the structure captured by the geometric approach is pretty simple, i.e., 3 or 4 links for most road junctions.

The topological approach is a holistic approach that considers the interrelationship of individual streets for better understanding the underlying structure of urban street networks. This is consistent with human cognition of large-scale environments. For example, in conveying a direction from location A to B, we often refer to a number of named streets that constitute the path. Using MSN Maps & Directions (mappoint.msn.com), we can easily make a query from where you live to where you work. The returned direction or path would consist of a number of named streets attached with mileages. It appears to suggest that topology is primary (at a structural level), and geometry is secondary (at a detailed level). In fact, a city is indeed an interconnected whole consisting of paths, edges, districts, nodes and landmarks at a cognitive level (Lynch 1960). In a similar fashion, a city can be perceived as an interconnected whole that links individual units (Alexander 1965). These individual units or constituents have different geometric sizes and shapes, but they are all perceived individually as a perceivable unit at a structural level. Alexander's (1965) "a city is not a tree" suggests that a city is neither a tree nor a regular lattice, but a semilattice with some shortcuts among some far-distant units. Interestingly enough, this semilattice structure is very similar to the structure that we found in the topologies of urban street networks in terms of street-street intersection.

The topologies of urban street networks have an efficient structure, which is demonstrated by a variety of other biological, technological, and social networks. The structure is called a small world network that is an intermediate status between an order and disorder. It has a higher efficiency in the sense of transporting information, goods or viruses at both local and global levels, when compared to their regular and random counterparts. This view of high efficiency can be seen from two prominent properties of small world networks, i.e. a small separation and a high degree of clustering. In a social setting, a small separation implies that any two arbitrarily chosen persons A and B are linked by a short chain of intermediate persons. The person A knows someone, who knows someone, … who knows the person B, and the number of someone is no more than 6. This is the so called "six degrees of separation", which has been empirically verified by the well known experiment conducted by Milgram (1967). It should be noted that the two persons A and B are randomly chosen from a large population, say, the entire population of the planet. On the other hand, small world networks tend to be highly clustered. It means that friends of a friend are likely to be friends. From a point of view of information flow, information can be diffused very quickly from one to another globally, and it also spreads efficiently among the circle of the friends locally. Although we use social networks as an example, the efficient structure explained holds true for a wide variety of real world networks (for comprehensive surveys, refer to Albert and Barabási 2002, Newman 2003, or more recently, Boccalettia et al. 2006, and Newman, Barabási and Watts 2006). Small world networks help to explain why computers can be infected overnight, and why the SARS virus can diffuse so quickly among the large population of human beings. The beauty of the small world networks lies in the fact that it retains an efficient structure inherited from their parents: regular and random networks (*cf.* section 2 for more explanations).

This paper aims to measure network's efficiency or characterize structural properties of urban street networks from the point of view of multidimensional chain of connectivity (or flow). This work is mainly inspired by the efficiency view of small world networks, in terms of how information flows among the nodes of a complex network. In this connection, Latora and Marchiori (2001) suggested a new formulation of the small world theory from the efficiency point of view. However, the formulation, actually current literature in small world networks, does not fully capture the structural complexity of real world networks. What has not drawn enough attention is multidimensional chain of connectivity (Degtiarev 2000) for characterising structure of real world networks. The notion of multidimensional chain of connectivity, which is referred to as flow in the context of this paper, is a fundamental concept of Q-analysis (Atkin 1974) - a computational language for structural description. To illustrate the concept of flow, let's see how an innovation, which involves multiple expertises, diffuses among an organization or from one person to another. Clearly the innovation would diffuse among the individuals who have the multiple expertises, NOT the individuals who lack the multiple expertises. Different innovations involve different multiple expertises. Thus an organization or a social network with the organization forms a multidimensional chain of connectivity (or flow for simplicity) for different innovations flow from one to another (*c.f.* section 3 for more details).

With the concept of flow, we define two relevant quantities: flow dimension and flow capacity, to measure the efficiency of individual nodes or that of an entire network. The concept and quantities are applied to three urban street networks to illustrate some nice structural properties. To our surprise for the topologies of urban street networks, previously confirmed as a form of small world and scale-free networks, we find that (1) the range of their flow dimension is rather wider than their random and regular counterparts, (2) their flow dimension shows a power-law distribution, and (3) they have a higher flow capacity than their random and regular counterparts. The findings have far reaching implications for understanding structure of urban street networks, as the topologies of which can be regarded a network infrastructure to service the flows of various kinds including vehicle, people, and goods.

The remainder of this paper is organized as follows. Section two introduces some basic mathematics of the small world and scale-free networks from a point of view of efficiency. Section three presents the concept and computation of flow using a simple notational street network. In section four, we report our major findings based on the computation of flow dimension and capacity for three urban street networks. Finally section five concludes the paper and points to our future work.

## 2. Mathematics of small world and scale-free networks

To introduce small world networks, let us start with the big world of a lattice illustrated in figure 1(a). Imagine that this regular lattice is expanded to include up to hundreds of millions of nodes, so that it becomes obviously a very big world. Generally speaking, the distance from any node to any other node tends to be very large, as in general one has to travel node by node to reach one's destination. However the situation can be significantly tipped, if we add some new edges that bring some far-distant nodes together (a sort of shortcuts) as illustrated in figure 1(b). The shortcuts significantly shorten the distance from any node to any other node in general, so the lattice is transferred to a semilattice – a small world. To understand why the semilattice is a small world, we must introduce some basic mathematics of graphs and networks.

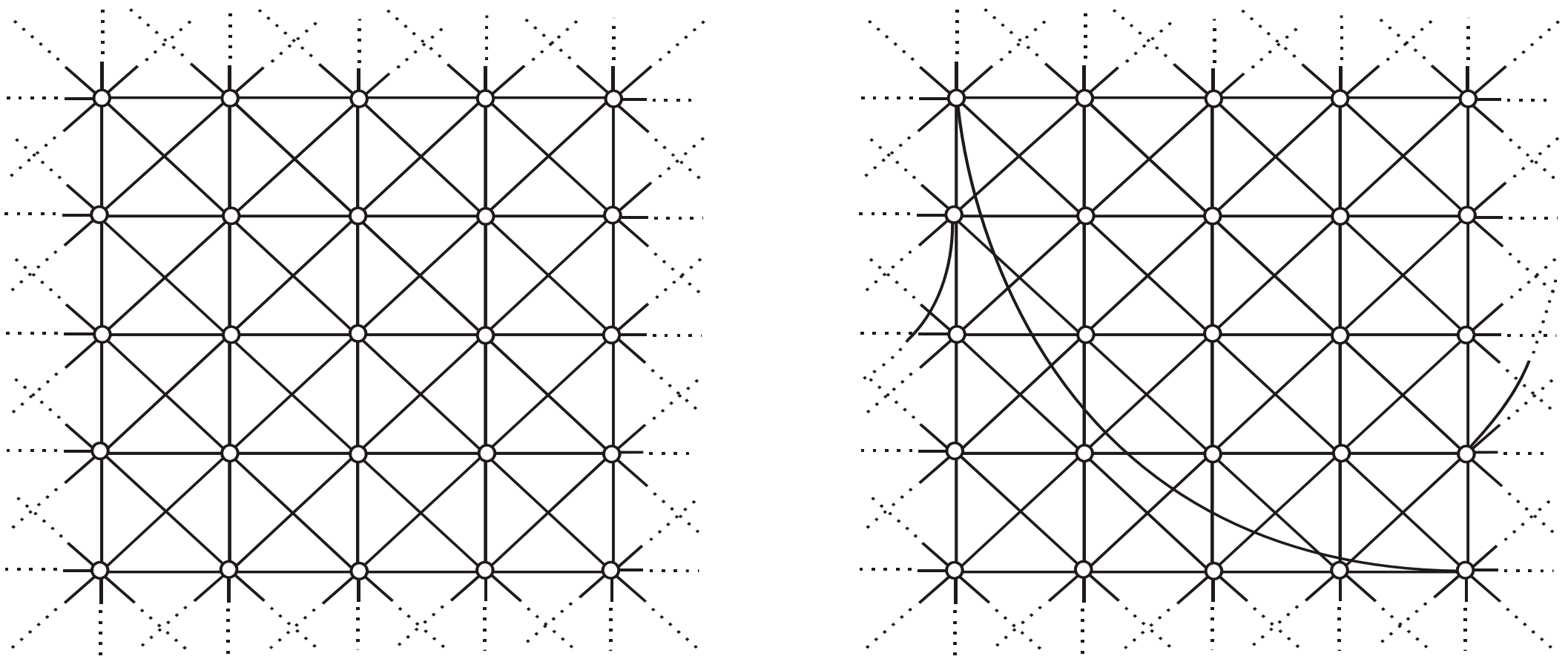

(a) (b)

Figure 1: Lattice (a) – a big world and semilattice (b) – a small world
(Note: different from the initial Watts' model (Watts and Strogatz 1998) and similar to Newman and Watts (1999), we randomly add a few new edges, but the new edges do not significantly change the average degree of the lattice)

A graph *G(V,E)* consists of a finite set of vertices (or nodes) $V = \{v_1, v_2, ...v_n\}$ (where the number of nodes is *n*) and a finite set of edges (or links) *E*, which is a subset of the Cartesian product $V \times V$ . The initial small world and scale-free networks are based on the connected graphs whose edges have neither directions nor weights. This kind of graph can be simply represented by an incidence matrix **R**(G), whose individual elements are defined by:

$$r_{ij} = \begin{cases} 1, & if\ v_i v_j \in E, \\ 0, & otherwise. \end{cases} \tag{1}$$

It is not difficult to note that the above incidence matrix **R**(G) is symmetric, i.e., $\forall r_{ij} \Rightarrow r_{ij} = r_{ji}$, and that all diagonal elements are equal to zero, i.e., $\forall r_{ij}|_{i=j} \Rightarrow r_{ij} = 0$. Thus the lower or upper triangular matrix of **R**(G) is sufficient for a complete description of the graph G. For a complete graph, in which every node links to every other, there are $n(n-1)/2$ edges. However, most real world networks are far sparser than a complete graph. If on average every node has *m* edges (average degree), then a sparse graph implies $m << n(n-1)/2$. The following two measures (or equivalently the initial small world and scale-free network models) are based on the undirected, unweighted, and connected graphs. However, this paper or definition of flow dimension and capacity has released the constraints. It implies that the concept of flow to be defined in what follows is applicable for directed, weighted and unconnected graphs as well.

Distance is a basic concept of graph theory (Buckley and Harary 1990), and fundamental for small world networks as well. The distance $d(i, j)$ between two vertices *i* and *j* of a graph is the minimum length of the paths that connect the two vertices, i.e., the length of a *graph geodesic*. For a given graph, the length of the maximum graph geodesic is called *graph diameter*. The distance of a given vertex $v_i$ far from all other vertices is called average path length. It is defined by

$$L(v_i) = \frac{1}{n-1}\sum_{j=1}^{n} d(i, j), \tag{2a}$$

The average ($\frac{1}{n}\sum_{i=1}^{n} L(v_i)$) of the average path length of the individual vertices is the average path length of the graph G,

$$L(G) = \frac{1}{n(n-1)} \sum_{i=1}^{n} \sum_{j=1}^{n} d(i, j), \qquad (2b)$$

From the equations, we might have a better sense of understanding of why the above lattice tends to be a big world. It is because that the average path length of a graph is increased with the square of the number of nodes ($\sqrt{n}$), i.e., the more nodes, and the longer the average path length. This is unlike the one-dimensional lattice whose path length is linearly increased. For a lattice that contains hundreds of millions of nodes, the average path length tends to be a large value. However, the introduction of some shortcuts into a large lattice can significantly reduce the average path length, and thus shrink the big world into a small one. Actually a random graph whose edges are randomly determined in terms of which node links to which node also has a short average path length according to random graph theory (Bollobás 1985), first studied by Erdös and Rényi (1960). For a random graph, the average path length is increased very slowly (logarithmically) with the number of nodes, i.e., $L_{random} = {}^{\ln(n)}/_{\ln(m)}$, where *m* is the average degree of the graph. So in terms of average path length, a small world network is very similar to a random network, i.e., a small separation between two arbitrarily chosen nodes of the network. This implies that a small world network has a very efficient structure for information flow at a global level.

Information flows efficiently not only at the global level, but also at a local level, i.e., in the circle of immediately neighbouring nodes. Again with the lattice in figure 1a, every node has eight neighbouring nodes. If the eight nodes all have an edge with each other (being a complete graph), then there would be in total $A_8^2 = 28$ possible edges. But actually there are only 12 edges as we can see. The ratio 12/28=0.43 indicates the degree of clustering for the eight neighbouring nodes, i.e., the higher the ratio, the more efficient information flows among the neighbours. Obviously the highest value of the ratio is 1, when the neighbours are highly connected as a complete graph, i.e. every node links to every other node. From the information flow point of view, the highly connected complete graph has a maximum efficiency. This can be seen from our daily lives. If all of my friends are friends with one another, then information about me can be spread maximally among the circle of my friends. In reality, friends of a person are hardly fully connected as a complete graph, BUT any two of the friends are *likely* to be friends in general. This implies a high clustering degree for a person or a social network in general. The clustering degree is measured by the clustering coefficient (Watts and Strogatz 1998) that can be simply defined by,

$$C(v_i) = \frac{\#\, of\;\; actual\ edges}{\#\, of\;\; possible\ edges}, \qquad (3a)$$

The average ($\frac{1}{n}\sum_{i=1}^{n} C(v_i)$) of the clustering coefficient of the individual vertices is the clustering coefficient of the graph G,

$$C(G) = \frac{1}{n} \sum_{i=1}^{n} \frac{\#\, of\;\; actual\ edges}{\#\, of\;\; possible\ edges}, \qquad (3b)$$

For an equivalent random graph (which means the same size and the same average degree of a random graph), its clustering coefficient is equal to the probability that two randomly selected nodes get connected, i.e., ${}^{m}/_{n}$. The ratio of *m* to *n* tends to a very small value (as a reminder, we are dealing with large sparse graphs). However, for an equivalent regular graph, its clustering coefficient tends to be a big value, given by $C_{lattice} = 0.43$. In summary, both the average path length and clustering coefficient can be good indicators for the efficiency of a real network respectively at the global and local levels, i.e. $E_{glob}(G)= 1/L(G)$, $E_{loc}(G)= C(G)$ (Latora and Marchiori 2001).

The high efficiency of small world networks comes from an important fact that for most of real networks like the Internet the connectivity is unevenly distributed among the nodes. In other words, connectivity is not randomly decided. A very few nodes are far more connected than the rest of others, i.e. vital few versus trivial majority. The highly connected nodes are called hubs. The special category of small world networks is called scale-free networks (Barabási and Albert 1999), which can be used to explain why a small world is a small world. It is because of the existence of hubs. It also helps to understand the nature of dynamics and evolution of many real world networks. It is important to note the relationship between the small world networks and scale-free networks. In general, a scale-free network is certainly a small world, but not vice verse. Small world networks can be put into different categories in terms of the nature of connectivity distribution (Amaral et al. 2000), and they all appear to have an efficient structure. In the next section, we will introduce the concept of flow for better characterizing the efficiency of networks or individual nodes.

## 3. Defining the concept of flow: dimension and capacity

Having introduced small world networks from the point of view of information flow, in this section we attempt to define the concept of flow. Flow is often referred to as movement of energy, material or information from one place to another through complex networks. In reality, there are different kinds of flows, such as vehicle flows, information flows, and flows of innovation, rumour, and virus. However, in this paper, we refer to flow as the multidimensional chain of connectivity, a structural property of networks or that of the individual nodes. How efficient a street network, or a street within the network, is in its capacity to accommodate/convey people or vehicle movement is subject to the underlying structure. The structure can be characterized by two flow related quantities: flow dimension and flow capacity.

For the sake of convenience and simplicity, but without loss of generality, we adopt a notional street network for illustration of the concept and computation of flow. The notional street network consists of 11 streets ($s_1$-$s_{11}$) and 20 junctions ($j_1$-$j_{20}$) (Figure 2). If we concentrate on the two pairs of streets <$s_1$, $s_4$> (in fact any pair from the set <$s_1$, $s_2$, $s_3$, $s_4$>, or equivalently any pair from the set <$s_5$, $s_6$, $s_7$, $s_8$>) and < $s_9$, $s_{10}$>, clearly the former pair has four other streets cutting across, while the latter has only two streets cutting across. It appears that the left lattice is more efficient than the right one in transporting people or vehicles. This observation provides a basic inspiration to develop the concept of flow for characterizing the structural property of a network, or that of individual nodes with the network.

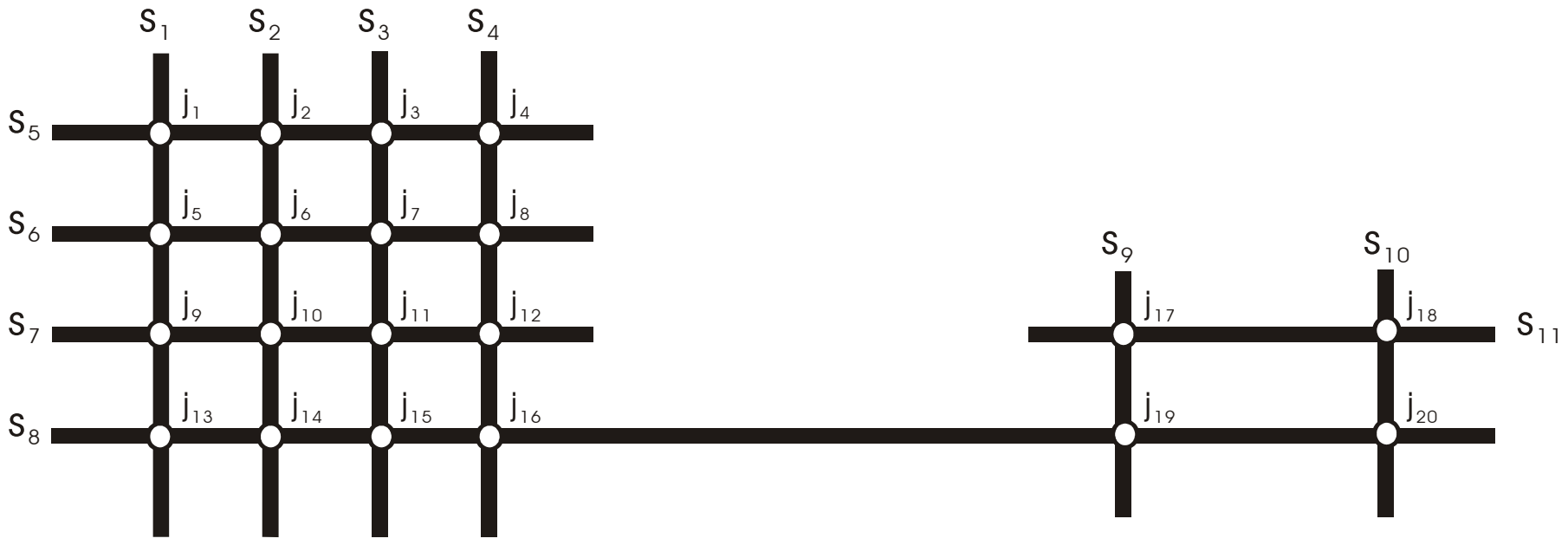

Figure 2: A notional street network

To further illustrate the concept of flow, we represent the notional street network topologically by taking individual streets ($s_1$ to $s_{11}$) as nodes and street intersections as links of a connected graph (Figure 3a). This connected graph is actually derived from a street-street incidence matrix $S$ (*cf.* matrix A1 in the appendix) (NOTE: the matrix S is still symmetric, but in more general, it can be asymmetric, representing a bigraph). To see how much flow there is between each pair of streets, we compute the flow matrix using the operation $FS = S * S'$ (*cf.* matrix A2 in the appendix). The flow matrix can be represented as a flow graph in figure 3b, where edge thickness represents flow dimensions. As a reminder, what *FS* represents is the number of liaison nodes (acting as a liaison role) between every pair of nodes. It should not be confused with the number of walks of length 2, which is obtained by $S * S$ , connecting nodes *i* and *j*, although they are same due to the symmetric nature of matrix *S* in this simple example. To this point, the reader may have found that the concept of flow is a bit "counter-intuitive", because the links in Figure 3a are not retained in Figure 3b. To explain this, it is important to note that what our model attempts to capture is the multidimensional chain of connectivity, rather than the simple connectivity represented in Figure 3a.

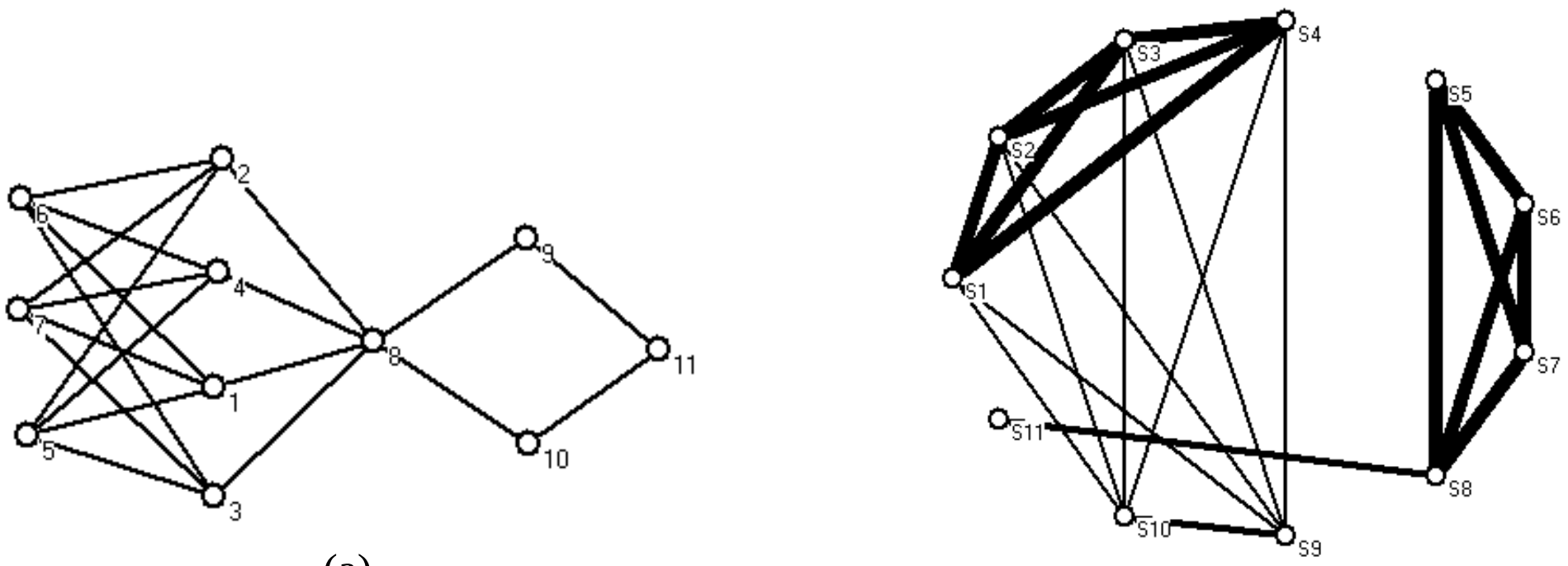

(a) (b)

Figure 3: The topology of the notional street network (a), and its flow graph (b)
(Note: flow dimension is indicated by edge thickness with respect to 1, 2, and 4 with the figure b)

The flow dimension refers to the number of liaison streets shared by a pair of streets, i.e., a sort of intersection strength between the pair. For instance, the pair <$s_1$, $s_4$> has four common streets (i.e., $s_5$, $s_6$, $s_7$ and $s_8$), while the pair <$s_9$, $s_{10}$> has just two liaison streets. The magnitude of flow dimension reflects the degree of flow between a pair of streets. Taking a path $s_2$-$s_3$-$s_9$ for example, the first half path ($s_2$-$s_3$) is very wide, while the second half path ($s_3$-$s_9$) very narrow. This flow graph is the starting point to define flow capacity. For the purpose of defining and computing the flow capacity, we slice the flow graph at different dimensional levels. First the flow graph itself represents the situation at dimension one, i.e, all edges above the dimension one are retained. We can filter out those edges whose flow dimension is less than one, and less than two, respectively. In the end, flow graphs at dimension two (figure 4a) and four (figure 4b) are formed. This slicing technique is taken from Q-analysis, and it is also given other names such as m-cores (Scott 1991), or m-slices (Nooy et al. 2005) in social network analysis.

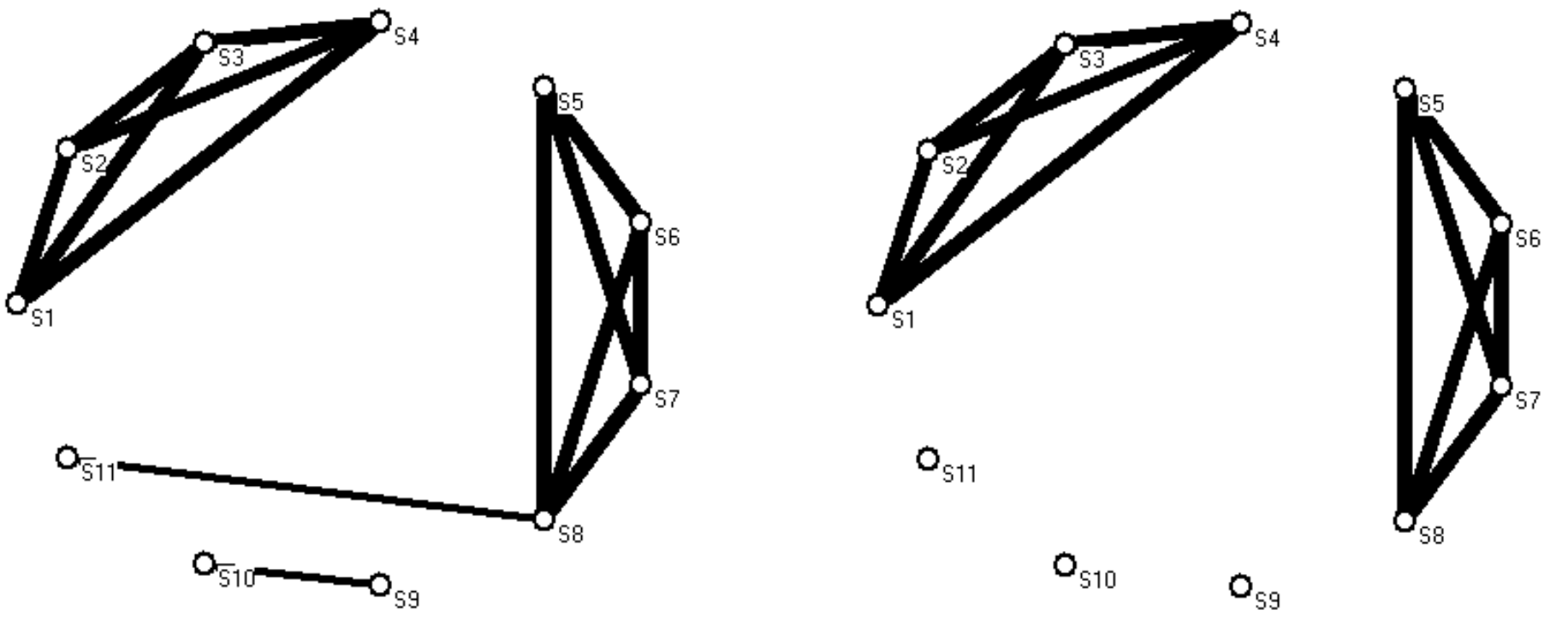

(a) (b)

Figure 4: Slicing the flow graph at the dimensions of two (a) and four (b)

Before formally defining the concept of flow capacity, we assume the sliced flow graphs serve as a network infrastructure for information flows. The flow graphs at lower dimensions can only transport low dimensional information, while the higher dimensional information needs higher flow graphs. This has an analogy to the different lanes of a street. For instance, bicycles can ride in a car lane, but cars cannot drive in a bicycle lane. The bicycles and cars are in analogy with lower dimensional and higher dimensional information in this case. Therefore the flow graph in figure 3(b) can only diffuse one-dimensional information flows, and the flow graph in figure 4(b) can diffuse four-dimensional information flows. Flow capacity is defined as a ratio of flow width to flow length. Taking the node $s_1$ for example, its flow capacity at dimension one with respect to figure 3b can be expressed by $\sum_{j=2}^{11} \frac{1}{d_1(1, j)} = 5$. With the expression, the numerator indicates dimension one and denominator represents the flow length (which is the shortest distance) from the particular node to all others. Clearly, the subscript of $d$ indicates dimension one. At dimension two and four (with respect to figure 4a and 4b), flow capacities are computed by $\sum_{j=2}^{11} \frac{2}{d_2(1, j)} = 6$ and $\sum_{j=2}^{11} \frac{4}{d_4(1, j)} = 12$. It should be noted that the flow graphs at dimension three and four are identical. So the flow capacity at dimension three is computed by $\sum_{j=2}^{11} \frac{3}{d_3(1,j)} = 9$ . The sum of the four flow capacities is the flow capacity of the node $s_1$, i.e., (5+6+9+12). To get rid of size effect, the sum is divided by 11*(11-1), so in the end the flow capacity of the node is (5+6+9+12)/(11*(11-1))=0.29

Having illustrated the computation of flow capacity with the example, we can remark that there are two factors that decide the flow capacity. One is called flow dimension and another is flow length (actually distance of geodesic at different dimensional levels). Intuitively, a higher flow dimension and a shorter flow length lead to a higher flow capacity. In general, for a vertex $v_i$ , its flow capacity is defined by

$$FC(v_i) = \frac{1}{(n-1)} \sum_{k=1}^{\ell} \sum_{j=1}^{n} \frac{k}{d_k(i,j)} \tag{4a}$$

where $\ell$ is the highest dimensional level.

The average of the flow capacity of the individual vertices is the flow capacity of the graph G,

$$FC(G) = \frac{1}{n(n-1)} \sum_{i=1}^{n} \sum_{k=1}^{\ell} \sum_{j=1}^{n} \frac{k}{d_k(i,j)} \tag{4b}$$

It is important to note that the above definition is illustrated with an undirected, unweighted and connected graph, but it holds true for a directed, weighted, and unconnected graph as well. Obviously a complete graph has a biggest flow capacity, since every node has a biggest flow dimension on the one hand, and a smallest flow length from every node to every other on the other. It should be noted that so far the flow capacity is defined as a measure for individual streets and the network of the streets. For the topological approach, both primal and dual approaches are transferable (Batty 2004, Jiang and Claramunt 2002). The above definition and computation can also be applied to the individual junctions of a street network. To achieve this, we must derive a street-junction incidence matrix first. Taking the notional street network for example, the street-junction matrix is represented as an incidence matrix of the size $11 \times 20$ , namely $SJ$ (see matrix A3 in the appendix). Actually the previous street-street incidence matrix $S$ can be derived using operation $S = SJ * SJ'$ (NOTE: the diagonal elements are set as zero, the same for the following matrix $J$ ). In the same fashion, we can derive the junction-junction incidence matrix $J$ , i.e., $J = SJ * SJ'$ (see matrix A4 in the appendix). Furthermore, using the junction-junction incidence matrix, the junction-junction flow matrix $FJ$ can be derived (see matrix A5 in the appendix). Based on the flow graph and using the above formulas, the flow capacity for the individual junctions and for the entire topology can be computed.

## 4. Computing the flow dimension and capacity for structuring urban street networks

We applied the concept and computation of flow to three urban street networks: Gävle (Sweden), Munich (Germany) and San Francisco (USA), including respectively 565, 785 and 637 streets – the same datasets used in our previous work (Jiang and Claramunt 2004). The reader may notice from figure 5 that Munich is a natural (self-organized or self-evolved) city, and San Francisco an artificial (or planned) city, whereas Gävle is somewhere between the natural and artificial cities, in particular with reference to the fact that two neighbours with Gävle are surrounded by ring roads. With the networks, we intend to illustrate some hidden structure or patterns, in comparison to their regular and random counterparts. A first step is to complete the transformation from a network to topology using the topological approach introduced at the beginning of the paper. Figure 5(d) illustrate the topology transformed from the San Francisco street network. A second step is to derive the flow graphs based on the topologies, and further compute the flow capacity for the individual nodes and for the entire topology. Three major findings are derived from the experiments.

The first finding is that the range of flow dimensions of the topologies is extremely wider than that of their regular and random counterparts (table 1). For instance, the Gävle topology involves 14 flow dimensions, while the number for its random and regular counterparts is only 2 and 3 (table 2). This fact can be further illustrated through a special visualization of the topologies. Again, taking the Gävle case for example, we arrange the topology layer by layer in terms of flow dimensions, i.e., from the bottom of a lowest dimension to the top of a highest dimension. Figure 5 demonstrates the fact of fourteen layers with the Gävle topology, and two layers with the random counterpart.

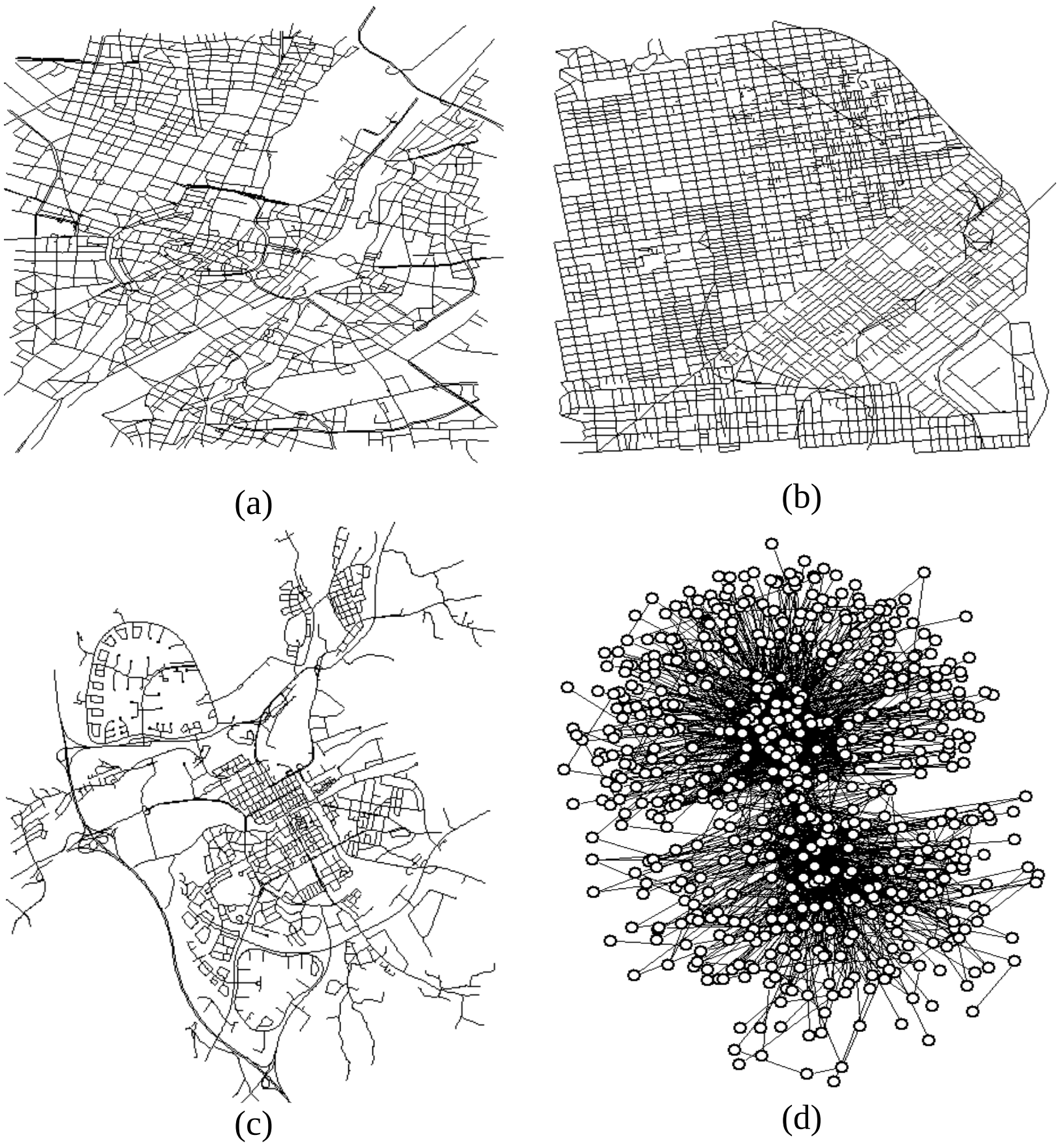

Figure 5: Street networks of Munich (a), San Francisco (b), Gävle (c) and the San Francisco topology (d)

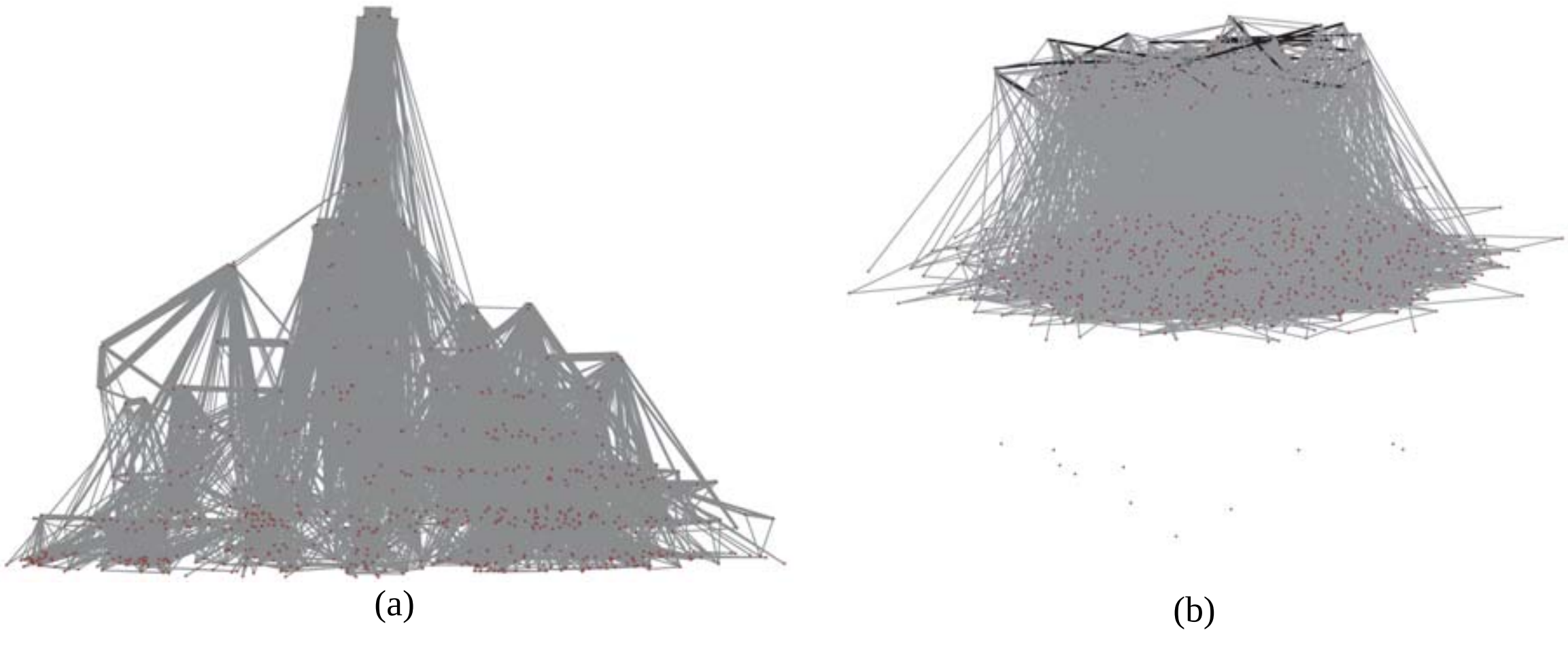

Figure 6: Flow graphs of the Gävle topology (a) and its random equivalent (b)

Table 1: The range of flow dimension for the three topologies and their counterparts
(RF = the range of flow dimension)

| | Gävle | Munich | San Francisco |
|---|---|---|---|
| RF | 14 | 11 | 46 |
| RF-rand | 2 | 2 | 3 |
| RF-reg | 3 | 4 | 7 |

The second finding is that the flow dimension demonstrates a power-law distribution with different connectivity exponents: 2.9 for Gävle, 3.6 for Munich, 1.2 for one part of San Francisco and 7.7 for another part (see figure 7 for the log-log plots). It implies that most streets have a quite low flow dimension, but a few streets have an extremely high flow dimension. This finding has a special implication for understanding traffic flows in urban systems. That is, always making those streets of an extremely high flow dimension through would significantly ease the traffic of an entire network system. It is important to note the two parts for the power-law tail with the San Francisco network, which is different from many others (Jiang 2007). The distributions of the flow dimension appear to fit in to that of the degree or street connectivity distribution illustrated by our previous work (Jiang and Claramunt 2004, Jiang 2005).

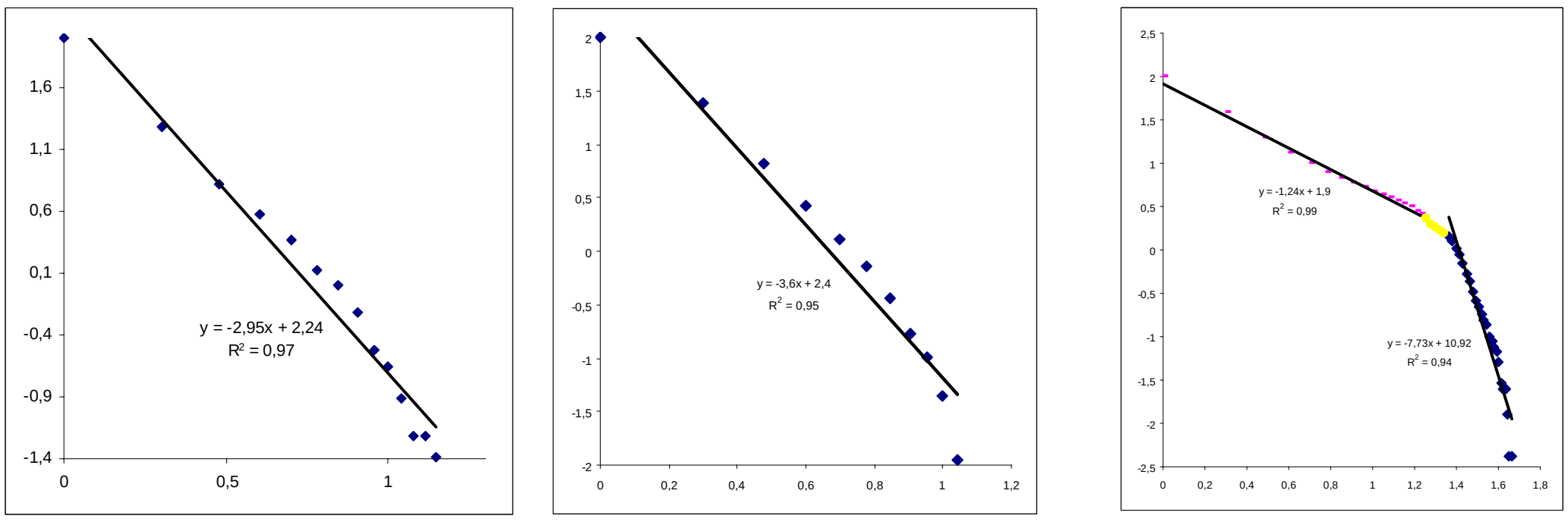

(a) (b) (c)

Figure 7: Cumulative distributions of flow dimension of Gävle (a), Munch (b) and San Francisco (c)
(The x-axes = flow dimension, and y-axes = cumulative probability)

The third finding is that the flow capacity of the topologies is larger than that of the random and regular counterparts (Table 2). This suggests that flow capacity can be a good indicator for determining whether or a real world network is a small world. This reminds us of the fact that the flow dimension resembles the clustering coefficient, and the flow length resembles the average path length. Therefore, the concept of flow capacity combines the initial two small world properties together into one formula, and it can be regarded as an alternative measure of network efficiency.

Table 2: Flow capacity (FC) of the three networks in comparison to their counterparts
(NOTE: n = size of networks, m = average degree)

| | Gävle | Munich | San Francisco |
|---|---|---|---|
| FC | 0.53 | 0.69 | 3.51 |
| FC-rand | 0.37 | 0.41 | 0.73 |
| FC-reg | 0.12 | 0.29 | 1.34 |
| m | 3.76 | 4.76 | 7.38 |
| n | 565 | 785 | 637 |

It should be noted that the above findings are made based on the primal topologies of urban street networks by taking the named streets as the nodes of the topologies, and eventually flow dimension and capacity are assigned to individual streets. However, we can also take the dual topologies of urban street networks by taking the junctions as the nodes of the topologies. This way the flow dimension and capacity can be computed and assigned to the individual junctions. Due to computational intensiveness, currently we are unable to obtain the flow capacity for the individual junctions of an entire network. However, our experiments based on partial networks illustrated the similar findings as above. Figure 8 illustrates the geographic distribution of the flow capacity for the individual streets and junctions of the Sätra neighbourhood network within Gävle. It is represented respectively as the line thickness and dot size. We can remark that the flow capacity for streets fits quite well to that of junctions. To this point, we have seen how the flow dimension and capacity are defined and computed for both streets and junctions of the urban street networks.

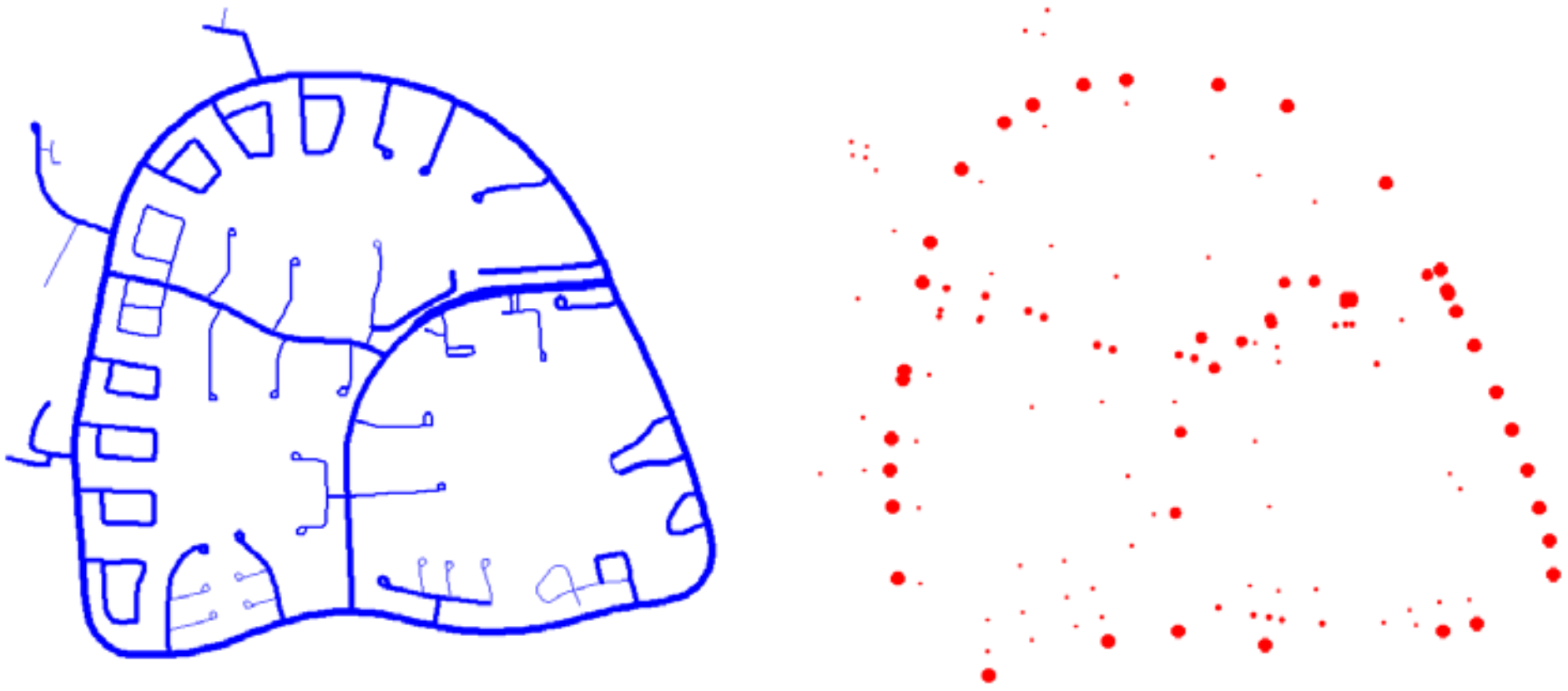

(a) (b)

Figure 8: Flow capacity for the individual streets (a) and junctions (b) of the Sätra neighbourhood

The above findings have some universal value. This is particularly true for knowledge or innovation diffusion which often needs multiple disciplinary expertises, in order for a knowledge chunk to be spread. For example, with an organization, individuals' expertise is different from one to another, and those who share expertises (common nodes) are likely to have a common language, and thus to have a knowledge chunk more quickly diffused from one to another. Those who have interdisciplinary background are likely to form m-cores for better knowledge diffusion. The concept of flow can help to differentiate from one organization network to another, or the individual persons' abilities in transferring knowledge within an organization.

## 5. Conclusion

This paper introduces the concept of multidimensional chain of connectivity or flow to characterize structural properties of urban street networks from both local and global levels. At the local level, it helps to differentiate between individual streets (or equivalently junctions) from one to another in terms of efficiency and it helps to identify key structural streets or junctions. At the global level, it further illustrates the fact that an urban street network departs it from its random and regular counterparts based on the single measure flow capacity. A major advantage of the flow capacity is that it takes multidimensional chain of connectivity into account while measuring efficiency of individual units or the entire network as a whole. In spite of the modelling capability, we must stress that the flow is a structural property, not the flows in reality such as vehicle and people flows. Whether the flow capacity can be a good indicator for real world flows is still an open question. It has implications for our future work.

Before any empirical verification of the concept and computation of flow, we tend to rely on its exploratory capability rather than possibilities of predicting real world flows. It helps to understand and explore traffic flows from a structural point of view. Taking the notional street network again for example, apparently it consists of two lattices: one to the left with four parallel streets at each direction crossed each other, and another to the right with two parallel streets crossed. From the structural point of view, the left lattice has a higher flow capacity than that of the right one. Thus it forms a gap or hole between the two lattices in terms of information flows. From the structural point of view, traffic jams are likely to occur between the two lattices. Of course, in reality traffic jams may not occur at all, if the number of vehicles is very low, or the vehicles are coordinated well. However the structural analysis based on the concept of flow capacity can help to indicate the possibility. We believe this is one of the major contributions of this paper besides the three major findings.

**Acknowledgements**
The author would like to thank Andrej Mrvar for his advice in using Pajek for partial computation and visualization with the study. The Munich dataset was provided by NavTech from the year 2000, the San Francisco dataset from ESRI sample data, and the Gävle dataset by Gävle city.

**Appendix: The matrixes associated with the notional street network**

The street-street incidence matrix S can be represented by,

$$S=\left[\begin{array}{c|ccccccccccc}
 & s_1 & s_2 & s_3 & s_4 & s_5 & s_6 & s_7 & s_8 & s_9 & s_{10} & s_{11} \\
s_1 & 0 & 0 & 0 & 0 & 1 & 1 & 1 & 1 & 0 & 0 & 0 \\
s_2 & 0 & 0 & 0 & 0 & 1 & 1 & 1 & 1 & 0 & 0 & 0 \\
s_3 & 0 & 0 & 0 & 0 & 1 & 1 & 1 & 1 & 0 & 0 & 0 \\
s_4 & 0 & 0 & 0 & 0 & 1 & 1 & 1 & 1 & 0 & 0 & 0 \\
s_5 & 1 & 1 & 1 & 1 & 0 & 0 & 0 & 0 & 0 & 0 & 0 \\
s_6 & 1 & 1 & 1 & 1 & 0 & 0 & 0 & 0 & 0 & 0 & 0 \\
s_7 & 1 & 1 & 1 & 1 & 0 & 0 & 0 & 0 & 0 & 0 & 0 \\
s_8 & 1 & 1 & 1 & 1 & 0 & 0 & 0 & 0 & 1 & 1 & 0 \\
s_9 & 0 & 0 & 0 & 0 & 0 & 0 & 0 & 1 & 0 & 0 & 1 \\
s_{10} & 0 & 0 & 0 & 0 & 0 & 0 & 0 & 1 & 0 & 0 & 1 \\
s_{11} & 0 & 0 & 0 & 0 & 0 & 0 & 0 & 0 & 1 & 1 & 0
\end{array}\right] \qquad \text{(A1)}$$

From the matrix S, we can derive its flow matrix $FS = S * S'$ as follows:

$$FS=\left[\begin{array}{c|ccccccccccc}
 & s_1 & s_2 & s_3 & s_4 & s_5 & s_6 & s_7 & s_8 & s_9 & s_{10} & s_{11} \\
s_1 & 0 & 4 & 4 & 4 & 0 & 0 & 0 & 0 & 1 & 1 & 0 \\
s_2 & 4 & 0 & 4 & 4 & 0 & 0 & 0 & 0 & 1 & 1 & 0 \\
s_3 & 4 & 4 & 0 & 4 & 0 & 0 & 0 & 0 & 1 & 1 & 0 \\
s_4 & 4 & 4 & 4 & 0 & 0 & 0 & 0 & 0 & 1 & 1 & 0 \\
s_5 & 0 & 0 & 0 & 0 & 0 & 4 & 4 & 4 & 0 & 0 & 0 \\
s_6 & 0 & 0 & 0 & 0 & 4 & 0 & 4 & 4 & 0 & 0 & 0 \\
s_7 & 0 & 0 & 0 & 0 & 4 & 4 & 0 & 4 & 0 & 0 & 0 \\
s_8 & 0 & 0 & 0 & 0 & 4 & 4 & 4 & 0 & 0 & 0 & 2 \\
s_9 & 1 & 1 & 1 & 1 & 0 & 0 & 0 & 0 & 0 & 2 & 0 \\
s_{10} & 1 & 1 & 1 & 1 & 0 & 0 & 0 & 0 & 2 & 0 & 0 \\
s_{11} & 0 & 0 & 0 & 0 & 0 & 0 & 0 & 2 & 0 & 0 & 0
\end{array}\right] \qquad \text{(A2)}$$

More generally, the notational street network can be represented by a street-junction incidence matrix $SJ$ :

$$SJ=\left[\begin{array}{c|cccccccccccccccccccc}
 & j_1 & j_2 & j_3 & j_4 & j_5 & j_6 & j_7 & j_8 & j_9 & j_{10} & j_{11} & j_{12} & j_{13} & j_{14} & j_{15} & j_{16} & j_{17} & j_{18} & j_{19} & j_{20} \\
s_1 & 1 & 0 & 0 & 0 & 1 & 0 & 0 & 0 & 1 & 0 & 0 & 0 & 1 & 0 & 0 & 0 & 0 & 0 & 0 & 0 \\
s_2 & 0 & 1 & 0 & 0 & 0 & 1 & 0 & 0 & 0 & 1 & 0 & 0 & 0 & 1 & 0 & 0 & 0 & 0 & 0 & 0 \\
s_3 & 0 & 0 & 1 & 0 & 0 & 0 & 1 & 0 & 0 & 0 & 1 & 0 & 0 & 0 & 1 & 0 & 0 & 0 & 0 & 0 \\
s_4 & 0 & 0 & 0 & 1 & 0 & 0 & 0 & 1 & 0 & 0 & 0 & 1 & 0 & 0 & 0 & 1 & 0 & 0 & 0 & 0 \\
s_5 & 1 & 1 & 1 & 1 & 0 & 0 & 0 & 0 & 0 & 0 & 0 & 0 & 0 & 0 & 0 & 0 & 0 & 0 & 0 & 0 \\
s_6 & 0 & 0 & 0 & 0 & 1 & 1 & 1 & 1 & 0 & 0 & 0 & 0 & 0 & 0 & 0 & 0 & 0 & 0 & 0 & 0 \\
s_7 & 0 & 0 & 0 & 0 & 0 & 0 & 0 & 0 & 1 & 1 & 1 & 1 & 0 & 0 & 0 & 0 & 0 & 0 & 0 & 0 \\
s_8 & 0 & 0 & 0 & 0 & 0 & 0 & 0 & 0 & 0 & 0 & 0 & 0 & 1 & 1 & 1 & 1 & 0 & 0 & 1 & 1 \\
s_9 & 0 & 0 & 0 & 0 & 0 & 0 & 0 & 0 & 0 & 0 & 0 & 0 & 0 & 0 & 0 & 0 & 1 & 0 & 1 & 0 \\
s_{10} & 0 & 0 & 0 & 0 & 0 & 0 & 0 & 0 & 0 & 0 & 0 & 0 & 0 & 0 & 0 & 0 & 0 & 1 & 0 & 1 \\
s_{11} & 0 & 0 & 0 & 0 & 0 & 0 & 0 & 0 & 0 & 0 & 0 & 0 & 0 & 0 & 0 & 0 & 1 & 1 & 0 & 0
\end{array}\right] \qquad \text{(A3)}$$

From this street-junction incidence matrix, we can easily derive the above street-street incidence matrix *S*, i.e. $S = SJ * SJ'$ (NOTE: diagonal elements of *S* are set to 0). In the same fashion, we derive a junction-junction incidence matrix $J = SJ * SJ'$ (NOTE: diagonal elements of *J* are set to 0) as follows:

$$
J = \left[\begin{array}{c|cccccccccccccccccccc}
 & j_1 & j_2 & j_3 & j_4 & j_5 & j_6 & j_7 & j_8 & j_9 & j_{10} & j_{11} & j_{12} & j_{13} & j_{14} & j_{15} & j_{16} & j_{17} & j_{18} & j_{19} & j_{20} \\
j_1 & 0 & 1 & 1 & 1 & 1 & 0 & 0 & 0 & 1 & 0 & 0 & 0 & 1 & 0 & 0 & 0 & 0 & 0 & 0 & 0 \\
j_2 & 1 & 0 & 1 & 1 & 0 & 1 & 0 & 0 & 0 & 1 & 0 & 0 & 0 & 1 & 0 & 0 & 0 & 0 & 0 & 0 \\
j_3 & 1 & 1 & 0 & 1 & 0 & 0 & 1 & 0 & 0 & 0 & 1 & 0 & 0 & 0 & 1 & 0 & 0 & 0 & 0 & 0 \\
j_4 & 1 & 1 & 1 & 0 & 0 & 0 & 0 & 1 & 0 & 0 & 0 & 1 & 0 & 0 & 0 & 1 & 0 & 0 & 0 & 0 \\
j_5 & 1 & 0 & 0 & 0 & 0 & 1 & 1 & 1 & 1 & 0 & 0 & 0 & 1 & 0 & 0 & 0 & 0 & 0 & 0 & 0 \\
j_6 & 0 & 1 & 0 & 0 & 1 & 0 & 1 & 1 & 0 & 1 & 0 & 0 & 0 & 1 & 0 & 0 & 0 & 0 & 0 & 0 \\
j_7 & 0 & 0 & 1 & 0 & 1 & 1 & 0 & 1 & 0 & 0 & 1 & 0 & 0 & 0 & 1 & 0 & 0 & 0 & 0 & 0 \\
j_8 & 0 & 0 & 0 & 1 & 1 & 1 & 1 & 0 & 0 & 0 & 0 & 1 & 0 & 0 & 0 & 1 & 0 & 0 & 0 & 0 \\
j_9 & 1 & 0 & 0 & 0 & 1 & 0 & 0 & 0 & 0 & 1 & 1 & 1 & 1 & 0 & 0 & 0 & 0 & 0 & 0 & 0 \\
j_{10} & 0 & 1 & 0 & 0 & 0 & 1 & 0 & 0 & 1 & 0 & 1 & 1 & 0 & 1 & 0 & 0 & 0 & 0 & 0 & 0 \\
j_{11} & 0 & 0 & 1 & 0 & 0 & 0 & 1 & 0 & 1 & 1 & 0 & 1 & 0 & 0 & 1 & 0 & 0 & 0 & 0 & 0 \\
j_{12} & 0 & 0 & 0 & 1 & 0 & 0 & 0 & 1 & 1 & 1 & 1 & 0 & 0 & 0 & 0 & 1 & 0 & 0 & 0 & 0 \\
j_{13} & 1 & 0 & 0 & 0 & 1 & 0 & 0 & 0 & 1 & 0 & 0 & 0 & 0 & 1 & 1 & 1 & 0 & 0 & 1 & 1 \\
j_{14} & 0 & 1 & 0 & 0 & 0 & 1 & 0 & 0 & 0 & 1 & 0 & 0 & 1 & 0 & 1 & 1 & 0 & 0 & 1 & 1 \\
j_{15} & 0 & 0 & 1 & 0 & 0 & 0 & 1 & 0 & 0 & 0 & 1 & 0 & 1 & 1 & 0 & 1 & 0 & 0 & 1 & 1 \\
j_{16} & 0 & 0 & 0 & 1 & 0 & 0 & 0 & 1 & 0 & 0 & 0 & 1 & 1 & 1 & 1 & 0 & 0 & 0 & 1 & 1 \\
j_{17} & 0 & 0 & 0 & 0 & 0 & 0 & 0 & 0 & 0 & 0 & 0 & 0 & 0 & 0 & 0 & 0 & 0 & 1 & 1 & 0 \\
j_{18} & 0 & 0 & 0 & 0 & 0 & 0 & 0 & 0 & 0 & 0 & 0 & 0 & 0 & 0 & 0 & 0 & 1 & 0 & 0 & 1 \\
j_{19} & 0 & 0 & 0 & 0 & 0 & 0 & 0 & 0 & 0 & 0 & 0 & 0 & 1 & 1 & 1 & 1 & 1 & 0 & 0 & 1 \\
j_{20} & 0 & 0 & 0 & 0 & 0 & 0 & 0 & 0 & 0 & 0 & 0 & 0 & 1 & 1 & 1 & 1 & 0 & 1 & 1 & 0
\end{array}\right] \qquad \text{(A4)}
$$

Based on the junction-junction incidence matrix, we can further derive a flow matrix using the operation of $FJ = J * J'$ (NOTE: diagonal elements of *FJ* are set to 0)

$$
FJ = \left[\begin{array}{c|cccccccccccccccccccc}
 & j_1 & j_2 & j_3 & j_4 & j_5 & j_6 & j_7 & j_8 & j_9 & j_{10} & j_{11} & j_{12} & j_{13} & j_{14} & j_{15} & j_{16} & j_{17} & j_{18} & j_{19} & j_{20} \\
j_1 & 0 & 2 & 2 & 2 & 2 & 2 & 2 & 2 & 2 & 2 & 2 & 2 & 2 & 2 & 2 & 2 & 0 & 0 & 1 & 1 \\
j_2 & 2 & 0 & 2 & 2 & 2 & 2 & 2 & 2 & 2 & 2 & 2 & 2 & 2 & 2 & 2 & 2 & 0 & 0 & 1 & 1 \\
j_3 & 2 & 2 & 0 & 2 & 2 & 2 & 2 & 2 & 2 & 2 & 2 & 2 & 2 & 2 & 2 & 2 & 0 & 0 & 1 & 1 \\
j_4 & 2 & 2 & 2 & 0 & 2 & 2 & 2 & 2 & 2 & 2 & 2 & 2 & 2 & 2 & 2 & 2 & 0 & 0 & 1 & 1 \\
j_5 & 2 & 2 & 2 & 2 & 0 & 2 & 2 & 2 & 2 & 2 & 2 & 2 & 2 & 2 & 2 & 2 & 0 & 0 & 1 & 1 \\
j_6 & 2 & 2 & 2 & 2 & 2 & 0 & 2 & 2 & 2 & 2 & 2 & 2 & 2 & 2 & 2 & 2 & 0 & 0 & 1 & 1 \\
j_7 & 2 & 2 & 2 & 2 & 2 & 2 & 0 & 2 & 2 & 2 & 2 & 2 & 2 & 2 & 2 & 2 & 0 & 0 & 1 & 1 \\
j_8 & 2 & 2 & 2 & 2 & 2 & 2 & 2 & 0 & 2 & 2 & 2 & 2 & 2 & 2 & 2 & 2 & 0 & 0 & 1 & 1 \\
j_9 & 2 & 2 & 2 & 2 & 2 & 2 & 2 & 2 & 0 & 2 & 2 & 2 & 2 & 2 & 2 & 2 & 0 & 0 & 1 & 1 \\
j_{10} & 2 & 2 & 2 & 2 & 2 & 2 & 2 & 2 & 2 & 0 & 2 & 2 & 2 & 2 & 2 & 2 & 0 & 0 & 1 & 1 \\
j_{11} & 2 & 2 & 2 & 2 & 2 & 2 & 2 & 2 & 2 & 2 & 0 & 2 & 2 & 2 & 2 & 2 & 0 & 0 & 1 & 1 \\
j_{12} & 2 & 2 & 2 & 2 & 2 & 2 & 2 & 2 & 2 & 2 & 2 & 0 & 2 & 2 & 2 & 2 & 0 & 0 & 1 & 1 \\
j_{13} & 2 & 2 & 2 & 2 & 2 & 2 & 2 & 2 & 2 & 2 & 2 & 2 & 0 & 4 & 4 & 4 & 1 & 1 & 4 & 4 \\
j_{14} & 2 & 2 & 2 & 2 & 2 & 2 & 2 & 2 & 2 & 2 & 2 & 2 & 4 & 0 & 4 & 4 & 1 & 1 & 4 & 4 \\
j_{15} & 2 & 2 & 2 & 2 & 2 & 2 & 2 & 2 & 2 & 2 & 2 & 2 & 4 & 4 & 0 & 4 & 1 & 1 & 4 & 4 \\
j_{16} & 2 & 2 & 2 & 2 & 2 & 2 & 2 & 2 & 2 & 2 & 2 & 2 & 4 & 4 & 4 & 0 & 1 & 1 & 4 & 4 \\
j_{17} & 0 & 0 & 0 & 0 & 0 & 0 & 0 & 0 & 0 & 0 & 0 & 0 & 1 & 1 & 1 & 1 & 0 & 0 & 0 & 2 \\
j_{18} & 0 & 0 & 0 & 0 & 0 & 0 & 0 & 0 & 0 & 0 & 0 & 0 & 1 & 1 & 1 & 1 & 0 & 0 & 2 & 0 \\
j_{19} & 1 & 1 & 1 & 1 & 1 & 1 & 1 & 1 & 1 & 1 & 1 & 1 & 4 & 4 & 4 & 4 & 0 & 2 & 0 & 4 \\
j_{20} & 1 & 1 & 1 & 1 & 1 & 1 & 1 & 1 & 1 & 1 & 1 & 1 & 4 & 4 & 4 & 4 & 2 & 0 & 4 & 0
\end{array}\right] \qquad \text{(A5)}
$$

We can remark that both flow matrixes (A5) and (A2) have the same range of flow dimension.